# Surface orbital order and chemical potential inhomogeneity of the iron-based superconductor FeTe$_{0.55}$Se$_{0.45}$ investigated with special STM tips

Dongfei Wang,[1,*] Ruidan Zhong,[2,†] Genda Gu,[2] and Roland Wiesendanger[1,‡]

[1]*Department of Physics, University of Hamburg, Jungiusstrasse 11, 20355, Hamburg, Germany*
[2]*Condensed Matter Physics and Materials Science Department, Brookhaven National Laboratory, Upton, NY 11973, USA*

The atomically clean surface of the iron-based superconductor FeTe$_{0.55}$Se$_{0.45}$ is investigated by low-temperature STM with different tip apex states. We show that a very sharp tip can enhance the scattering channel between the Γ and X/Y points of the sample's Brillouin zone. Furthermore, by manipulating a single Fe atom onto the tip apex, signatures of the orbital nature of the subsurface Fe layer of FeTe$_{0.55}$Se$_{0.45}$ can be identified. By preparing a charged tip state, the intrinsic spatial inhomogeneity of the chemical potential of FeTe$_{0.55}$Se$_{0.45}$ can be revealed. As a result, three different types of vortex bound states originating from locally varying topological properties of the FeTe$_{0.55}$Se$_{0.45}$ surface are observed by scanning tunneling spectroscopy.

During the last decades, a lot of efforts have been devoted to the functionalization of scanning tunneling microscope (STM) probe tips. For instance, to make the tip apex magnetically sensitive, electrochemically etched tips made of bulk magnetic materials like $CrO_2$ (*1*), Cr (*2*), Ni (*3, 4*) and Co (*5*) were used. Nonmagnetic tips covered by an ultrathin film (*6, 7*) of magnetic materials were also developed for reducing the tip's stray field. Carefully etched Ag(*8*) and Au(*9*) bulk tips with a sharp and smooth apex were employed in the study of tip-enhanced Raman spectroscopy and light emission (*10*). By decorating the tip apex with a single molecule (*11*), one can even get information about the chemical bonding (*12, 13*), exchange (*14*) and superexchange (*15*) interactions of molecules on metal surfaces.

Special functionalized tips were also used for STM and scanning tunneling spectroscopy (STS) studies of superconductors. To achieve higher energy resolution, tips made of superconducting materials were employed (*16*). Later on, the application of superconducting probe tips was extended to the study of fundamental properties of superconductors (*17, 18*), the transport characteristics of Josephson junctions (*19-21*) and recently the pair density wave state (*22-24*). However, most of the studies were focused on conventional superconductors and cuprates. Investigations of iron-based superconductors with special functionalized STM tips are still rare (*25, 26*).

Here, we employ three different kinds of special STM tips to study the iron-based superconductor $FeTe_{0.55}Se_{0.45}$. First, by making use of a very sharp tip, we have found a significant enhancement of the quasiparticle scattering from the Γ point to the X/Y point compared to a normal tip. Second, by picking up a single Fe atom from the sample surface to the tip apex, we have successfully observed a new type of order on the $FeTe_{0.55}Se_{0.45}$ surface. We attribute these patterns to the sensitivity of a special tip state to the Fe d-orbital ordering. Third, we can even image the charge or chemical potential distribution of the $FeTe_{0.55}Se_{0.45}$ surface with a charged STM tip. The spatial variations of the chemical potential offer an explanation for the observation of three different types of superconducting vortices hosting different kinds of vortex bound states, namely topologically trivial Caroli-de Gennes-Matricon or topologically non-trivial Majorana bound states.

The STM/STS measurements were performed in ultra-high vacuum (less than $10^{-10}$ mbar) and at a temperature of 1.1 K. A lock-in technique with a bias modulation of 0.03 mV and a frequency of 893 Hz was used to record differential tunneling conductance (dI/dV) spectra. Before recording the spectra, the STM tip has been stabilized with a sample bias voltage

$V_{stab}$ and a tunneling current $I_{stab}$. During the acquisition of dI/dV spectra, the STM feedback loop has been switched off. All the data have been obtained with a bulk Cr tip as a starting point which then has been modified by various in-situ tip treatment procedures as explained in the following.

Previous studies showed that the spatial resolution of STM images can significantly be improved by picking up a single atom from a surface onto the STM tip (*27, 28*). Here, we repeatedly transferred individual Fe atoms onto the tip by vertical manipulation of pre-deposited Fe atoms on the surface of $FeTe_{0.55}Se_{0.45}$. Atomic-resolution STM images of the sample surface before and after the tip has become extremely sharp are shown in Fig. 1(a) and (e). The two bright protrusions are Fe adatoms with the left one partially buried into the surface. Fig. 1(e) clearly shows a reduction of the apparent size of the individual Fe atoms, while the Te/Se square lattice is much more clearly visible compared to Fig. 1(a). The enhanced atomic-scale contrast is also observed in the corresponding FFT maps of Fig. 1(b) and (f): additional second-order peaks can be seen in Fig. 1(f) with the sharper tip as compared to (b). Furthermore, the local electronic density of states (LDOS) distribution can be visualized more clearly with the very sharp tip as becomes obvious by comparing Figs. 1(c) and (g). The quasiparticle interference (QPI) pattern obtained at 1 mV with a normal tip mainly highlights the LDOS variations as shown in Fig. 3(c), while with the very sharp tip, the QPI image at 1 mV also reveals the atomic lattice very clearly as shown in (g). These differences can even more clearly be seen by comparing the corresponding FFT maps in Fig. 1(d) and (h). The X/Y spot intensities in Fig. 1(h) are greatly enhanced compared to those in (d). Previously, such intensity enhancement has been observed in experiments with a high external magnetic field and was attributed to the non-conventional pairing symmetry of iron-based superconductors (*29-31*). However, in our case, the difference in intensity of the Fourier spots can be explained by a trivial tip sharpening effect.

By transferring single Fe atoms onto our tip apex, it is not only the spatial resolution which can be greatly improved, but the tip can additionally become sensitive to the surface Fe orbital ordering. This is supported by a newly observed stripe pattern on the surface of $FeTe_{0.55}Se_{0.45}$ other than the normal Te/Se square lattice, as shown in Figure 2. Figure 2(a) and (b) display STM images of the same surface region obtained with a normal tip and a special tip having picked-up Fe atoms at its apex. One can clearly see a new periodic stripe pattern in the off-diagonal Te/Se lattice direction. These patterns can be observed everywhere on the surface: the STM image of another region imaged with the same tip is shown in Fig. 2(c,e,f). The green arrow k in Fig. 2(c) indicates the propagation direction of the periodic stripe pattern which is the same as in Fig. 2(b). This additional periodicity can even more clearly be identified in the corresponding FFT map of Fig. 2(c), as shown in Fig. 2(d). Besides the X/X' and Y/Y' spots which reflect the square atomic lattice symmetry, new spots K and K'(-K) appear. By carefully analyzing the FFT data we found that the angle between ΓX and ΓK is 45° and that the length ratio of the corresponding wavevectors is $1/1.43 \sim 1/\sqrt{2}$. These values correspond in fact to the Fe square lattice underneath the top layer Te/Se lattice. As shown in Fig. 2(g), the Fe atoms reside at the bridge sites of the top layer Te/Se lattice with a nearest neighbor distance of $1/\sqrt{2}$ the Te/Se lattice constant. Furthermore, the atomic Fe lattice is rotated by 45° relative to the surface Te/Se lattice. Thus, we attribute the newly observed stripe pattern to the Fe lattice below the surface (Fig. 2(h)). However, it still needs to be investigated further which information was obtained by the special tip. It should also be noted that the brighter regions which are mainly composed of surface Te atoms always exhibit the square lattice symmetry (*32*).

A previous study showed that the spin-polarized Yu-Shiba-Rusinov states of individual Fe atoms on the $FeTe_{0.55}Se_{0.45}$ surface can be revealed by using a bulk Cr tip with an Fe atom at its apex (*26*). Therefore, our tip might be spin-polarized as well. It is known that FeTe, which is quite similar to $FeTe_{0.55}Se_{0.45}$, exhibits a bi-collinear antiferromagnetic order at its surface (*33, 34*). However, our observation here differs from the FeTe case in two ways: First, the period found for the FeTe surface measured with a spin-polarized tip is twice the Te lattice constant whereas in our case we observe a period of $1/\sqrt{2}$ the Te lattice constant for the newly observed pattern. More importantly, the direction of the stripe pattern observed on the FeTe surface is the same as for the Te lattice, in contrast to the new pattern observed in the present study. Thus, this new pattern is not related to the bi-collinear unidirectional stripe order observed on FeTe surfaces with a spin-polarized tip. We even confirmed that the newly observed pattern is not related with magnetic order by performing magnetic-field dependent STM measurements. As shown in Figure S1, we can hardly see any contrast change when we ramp the externally applied magnetic field between 2T and -2T.

We therefore propose another type of explanation: The stripe pattern we observe might be related with the hybrid Fe d-orbitals probed by the d-orbital of the Fe atom at the tip apex. This interpretation is supported by two experimental facts: 1) The stripe pattern propagates both in the diagonal and the off-diagonal direction of the Te/Se lattice, depending on the particular tip preparation. 2) The new stripe pattern always appears in the darker region of the $FeTe_{0.55}Se_{0.45}$ surface where mainly Se atoms are concentrated (*32*). Previous reports showed that near the Γ point, the top valence bands include the Fe-Fe $d_{yz}$, $d_{xz}$ bonding states with finite energy splitting, and the $d_{xy}$ states in the middle (*35, 36*). The $d_{xz}$ band and the $d_{xy}/p_z$ hybrid band $\Gamma_2^-$ have more $p_z$ orbital component of the chalcogen atoms. Let us first assume the electrons tunnel from the

$d_{yz}$ orbital of the Fe atom tip with energy close to the Fermi level where $d_{yz}/d_{xz}$ orbital states dominate. We can expect that the tunneling probability between the tip's $d_{yz}$ orbital and the Fe $d_{yz}$ bonding orbital is sufficiently large while the tunneling between the tip's $d_{yz}$ orbital and the Fe $d_{xz}$ orbital is low. This is demonstrated by Fig. 2(h). Therefore, we can see the stripe period only in one direction for some special tips. It also explains why the newly observed stripe pattern only appears in the darker regions, as the $d_{yz}$ orbital is less screened by the chalcogenide $p_z$ orbital. We can also understand why the tip preparation success rate for this kind of contrast is so low because the tip apex Fe atom need to achieve a special configuration where only $d_{yz}$- or $d_{xz}$-orbitals are involved in the tunneling. This particular configuration is rather fragile and a small bias pulse can change it. We further verify our interpretation by performing current- and bias voltage-dependent measurements (Figs. 2(e,f)): We find that the stripe pattern is indeed sensitive to the sample bias voltage as can be seen in Figure 2(f). The particular stripe pattern is clearly resolved at -10 mV but disappears at -40mV. This is consistent with the fact that when imaging the surface with larger bias, more electrons from the bands including the $p_z$ orbital states of the chalcogenide atoms are involved in the tunneling process (*35, 36*). On the other hand, the stripe pattern is not sensitive to the tunneling current as shown in Figure 2(e). Our interpretation is also consistent with our magnetic field dependent measurements because a pure orbital imaging mechanism does not involve contributions from spin-polarized tunneling. These findings provide novel microscopic insights into the nature of orbital order in iron-based superconductors (*37, 38*).

In a next series of studies, we were picking up some larger clusters of FeTeSe material from the surface by the STM tip, thereby allowing the imaging of sample specific inhomogeneities on the nanometer scale. This new observation is documented by Figs. 3(a)-3(d) which represent STM images being recorded with the same sample bias voltage of -10 mV, but different tunneling currents of 50 pA, 100 pA, 200 pA and 400 pA, correspondingly. Most notably, some bright regions with sharp boundaries appear in those constant-current STM images. The spatial extensions of the bright areas diminish with increasing tunneling current. Both the Te/Se atomic square lattice and the nano-scale bright regions can be imaged at the same time. The observed bright contrast features remind us about the study of charging effects of dopants in semiconductors (*39-43*) and molecules (*44, 45*). In those cases, the dopant energy level is bent by the tip-induced electric field and results in a controllable charged and uncharged state of the dopant depending on the tip's lateral position. This effect manifests itself by the observation of a circular feature around the dopant whose diameter depends on the sample bias used for differential tunneling conductance mapping. However, in our case, the observed bright features differ in three ways: 1) they are not ring-like, but can have arbitrary shapes and do not necessarily have a dopant in the center. 2) The shapes of the bright features are not significantly affected by using a larger negative bias voltage for scanning as shown in Figure 3(e-g). 3) The STM tips used for the previous studies of dopant charging and discharging were normal metal tips. However, the new surface inhomogeneous phase was never reported for FeTeSe samples studied with W- or Cr-tips using similar tunneling conditions (*26, 29, 46, 47*). Considering all these experimental observations, we propose a new charge imaging mechanism: The Te/Se surface exhibits nano-scale charge and chemical potential variations and the resulting nano-scale inhomogeneities can be imaged by a semiconducting tip with dopants. As illustrated in Figs. 4(a,b), for a metal tip with such a doped semiconducting cluster at its apex, the energy level of the dopant's ionic state $E_D$ can be located quite close to the conduction band but below the Fermi energy of the tip. When the STM tip is located above a normal area of the FeTeSe sample, only states of the conduction band of the tip are involved in the tunneling. However, when the tip is located above a charged area as shown in Figure 4(c), this ionic state will be lifted due to the locally induced electric field if the area is negatively charged. When the electric field is strong enough to lift the ionic state above the conduction band, the dopant will lose its electrons to the conduction band and will become charged. The electric field of the ionized dopant will in turn lower the energy of the conduction band of the tip (*40*). Thus, additional electron tunneling occurs from the charged area and gives rise to an increase in apparent topographic height in constant-current STM scans. When we move the tip closer to the sample as shown in Figs. 3(a-d), the tip experiences a larger negative electric field and also the negative sample bias applied may induce some additional band bending. As a result, the conduction band will be lifted again due to a higher electronic potential resulting in a reduction of tunneling events in the charged areas compared to the surroundings. Therefore, the spatial extension of the charged area decreases rapidly with decreasing tip-sample distance. In the bias-dependent experiment shown in Figs. 3(e-h), the observations can be interpreted as follows. A higher sample bias voltage will lead to two effects: on one hand, the tip-sample distance will increase in the constant-current STM mode and the electric field will decrease correspondingly. On the other hand, the electric field becomes stronger due to a larger sample bias. Therefore, the experimental results shown in Figs. 3(e-h) could be a competing result of those two effects, which finally result in an only moderate change of the spatial extensions and shapes of the charged areas. Spatially inhomogeneous charge distributions of topological materials have also been observed recently by Edmonds et al. for a $Na_3Bi$ sample (*48*).

Our assumption of a spatially inhomogeneous charge and chemical potential distribution is consistent with the observation that not every vortex core of superconducting FeTe$_{0.55}$Se$_{0.45}$ in an external magnetic field shows a Majorana zero mode (*46*). One possible explanation is that the chemical potential in the material exhibits a spatially inhomogeneous distribution (*49*). The relative positions of the Fermi energy and the Dirac point could result in three different topological phases of the FeTe$_{0.55}$Se$_{0.45}$ surface (*50*). When the Fermi energy is located well below the topological insulator's Dirac cone, bulk bands play the most important role and the vortex core exhibits Caroli-de Gennes-Matricon bound states (*51*). In contrast, a Majorana zero mode (MZM) appears within the vortex core when the Fermi energy is located in the vicinity of the topological Dirac cone. By further bringing the Fermi energy up to the vicinity of the topological semimetal Dirac cone, the surface will exhibit a topological semimetal phase with a helical Majorana mode inside the vortex line (*52*). Interestingly, we observed three different kinds of bound states within the vortices of our superconducting FeTe$_{0.55}$Se$_{0.45}$ sample, as shown in Fig. 5. The vortex showing a MZM is presented in Fig. 5(a), and a corresponding STS line-cut is shown in Fig. 5(b). From the set of individual tunneling spectra as a function of spatial position, we can clearly see a zero-energy mode inside the superconducting gap. In contrast, a vortex exhibiting a Caroli-de Gennes-Matricon bound state (CBS) is presented in Fig. 5(c). From the corresponding line-cut displayed in Fig. 5(d) we can identify a bound state at 0.28mV being located near the center of the vortex core and two symmetric peaks at +/- 0.79 mV located outside of the vortex core. If we identify the peak at 0.28 mV as the first CBS and the one at 0.79 mV as second CBS, then we find a difference in energy of 0.51 meV between these two states. This value is almost twice the energy of the first CBS. Thus, both the energy values and the spatial distribution of the peaks fit very well to theory (*51*). On the other hand, we observe a new kind of vortex with a zero energy bound state at the boundary and a clearly off-zero peak in the center of the vortex core. The STS map and the corresponding line-cut are shown in Figs. 5(e,f). Based on the assumption that some parts of our sample surface are negatively charged, the observation of this new type of vortex can be explained straightforwardly. According to theory, a helical Majorana mode exists in the bulk vortex line. However, at the surface, where the vortex line terminates, a bound state close to the Fermi energy should appear according to the bulk-boundary correspondence. Thus, we can expect a splitting of this bound state at the vortex core due to a significant hybridization, but less splitting at the vortex boundary. Our analysis fits the experimental data shown in Figure 5(f) qualitatively. For a quantitative analysis, further experiments and theoretical studies are required.

In conclusion, by functionalizing the STM tip apex by a single Fe atom or a cluster of

FeTe$_{0.55}$Se$_{0.45}$ material, we succeeded in the observation of several new phenomena for the iron-based superconductor FeTe$_{0.55}$Se$_{0.45}$. The scattering channel between the Γ and the X/Y points can be greatly enhanced by an atomically sharp STM tip. Based on electron tunneling from a special orbital of the Fe-atom tip, we imaged the dxz/dyz orbital order of the Fe sublattice in FeTe$_{0.55}$Se$_{0.45}$. A semiconducting tip with a dopant atom allowed us to resolve the surface charge distribution inhomogeneity which can explain the occurrence of different types of bound states observed in three different vortices of superconducting FeTe$_{0.55}$Se$_{0.45}$.

## ACKNOWLEDGEMENT

We would like to thank Jens Wiebe, Thore Posske, Ching-Kai Chiu, De-Liang Bao, and Lingyuan Kong for useful discussions as well as Torben Hänke and Anand Kamlapure for technical support. We also thank Hong Ding, Fazhi Yang and Cuihua Liu for providing the samples. This work has been supported by the EU via the ERC Advanced Grant ADMIRE (No. 786020), the DFG via the Cluster of Excellence "Advanced Imaging of Matter" (EXC 2056, project ID 390715994) and the US Department of Energy, office of Basic Energy Sciences (contract no. de- sc0012704).

[*] dwang@physnet.uni-hamburg.de
[†] Present address: Tsung-Dao Lee Institute & School of Physics and Astronomy, Shanghai Jiao Tong University, Shanghai 200240, China
[‡] wiesendanger@physnet.uni-hamburg.de

Figure 1

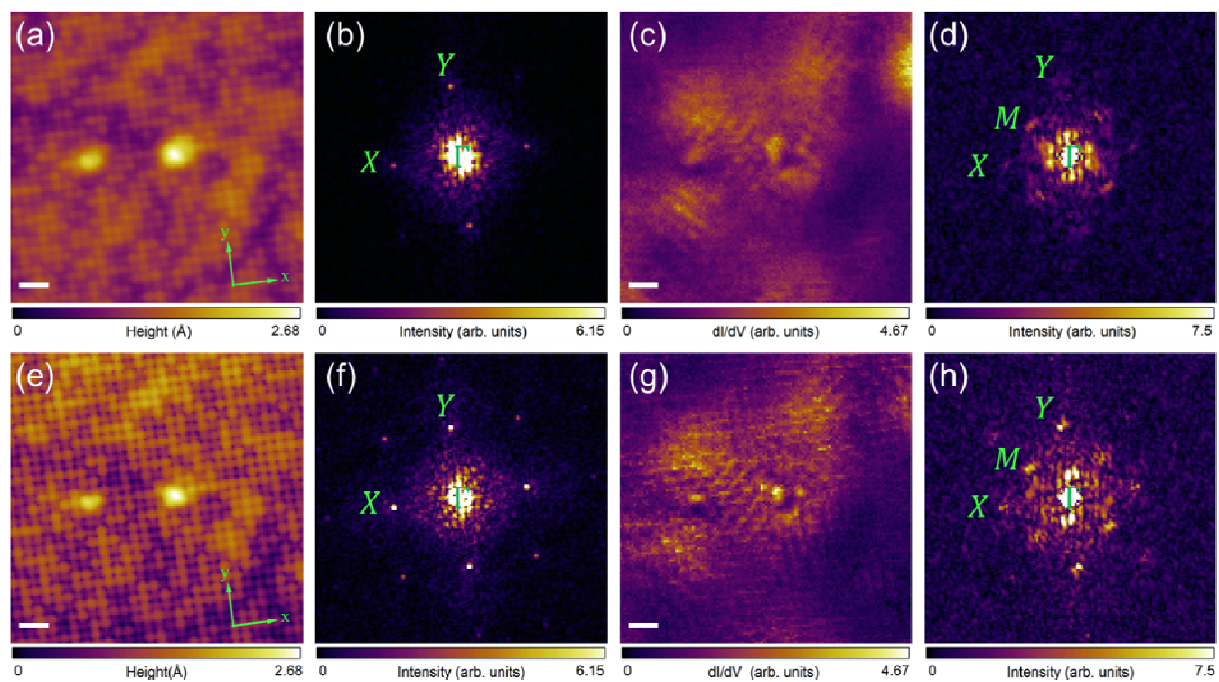

**Figure 1:** Atomic-resolution STM images and corresponding FFT maps before and after tip modification. (a,e) Constant-current STM topography in the vicinity of two Fe impurities before (a) and after (e) tip modification. Scale bar: 1 nm. (b,f) FFT maps of (a,e). and denote the first-order spots of the reciprocal lattice. The second-order spots can only be seen clearly in (f). (c,g) dI/dV maps taken at 1 mV for the same regions as shown in (a,e) before (c) and after (g) tip modification. A quasiparticle interference pattern can be seen in both (c) and (g). Scale bar: 1 nm. (d,h) FFT maps of (c,g). After tip modification, the intensity of the spots at points and is greatly enhanced in (h) compared to that in (d) before the tip modification. Tunneling parameters in (a,c,e,g): $V_{stab}$=-10 mV, $I_{stab}$=600 pA, $V_{osc}$= 0.03 mV.

Figure 2

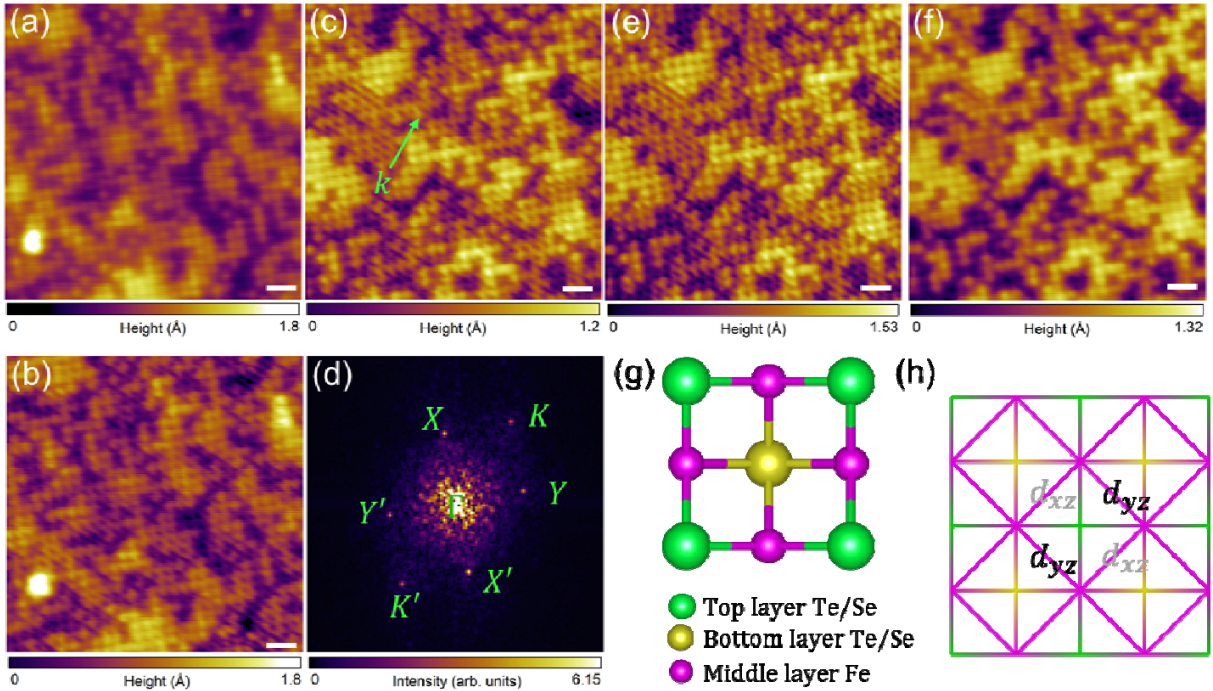

**Figure 2:** Imaging Fe orbital information with a special STM tip. (a-b) Atomically resolved constant-current STM topography image before (a) and after (b) the tip has been modified by the adsorption of a single Fe atom at its apex. A stripe pattern coexisting with the square lattice of the Te/Se lattice is visible in (b). (c) Atomic-resolution constant-current STM topography image of another area obtained with the same tip as used in (b). The direction of a special periodic pattern is highlighted by a green arrow labeled with k. Tunneling parameters in (a-c): V=-10 mV, I=400 pA. (d) FFT map of (c). New peaks labeled   and  ′ appear beside ±X and ±Y which reflect the Te/Se square lattice. (e) STM image of the same area as shown in (c) obtained with the same sample bias of -10 mV, but different tunneling current 1 nA. (f) STM image of the same area as shown in (e) with the same tunneling current 1 nA but a different sample bias of -40 mV. (g) Top-view model of the Fe (Te,Se) crystal. (h) Bonds between Te/Se and Fe atoms. Green/yellow corners and intersections represent the top/bottom Te/Se atoms. Purple corners and intersections represent the interlayer Fe atoms. Fe-Fe bonds are shown as purple lines between the Fe atoms. Scale bar in (a-c), (e,f): 1 nm.

Figure 3

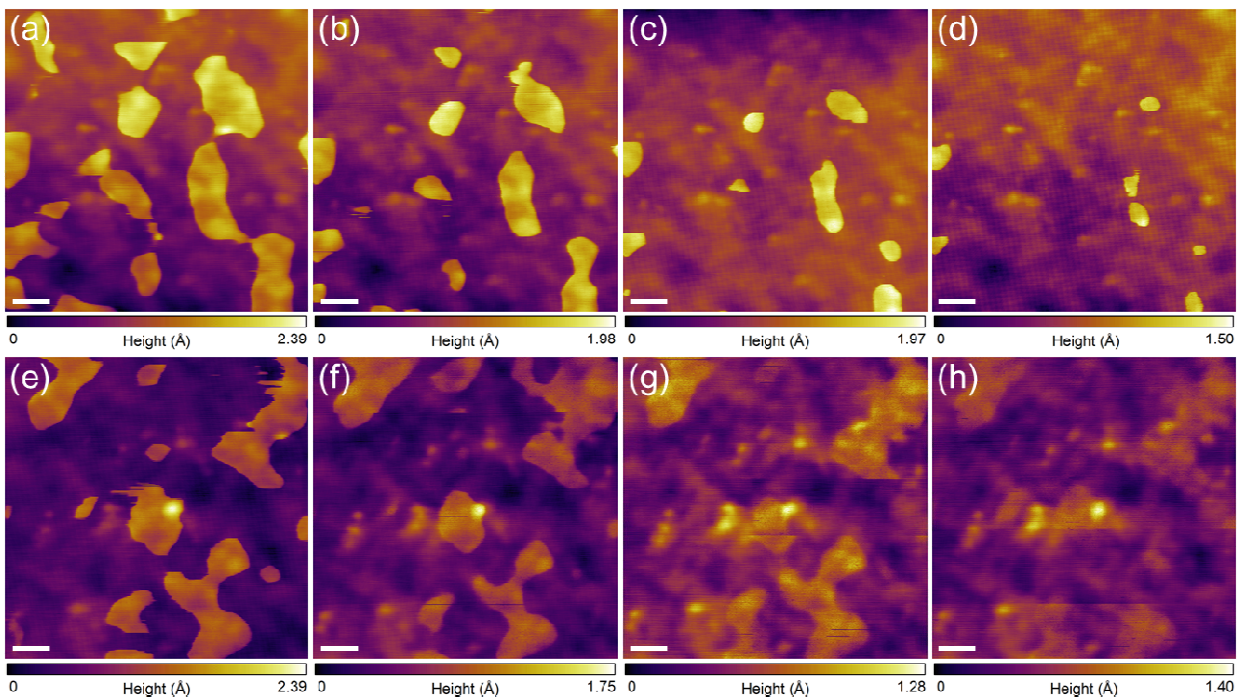

**Figure 3:** Charging effects revealed by a special STM tip. (a-d) Constant-current STM images of the same region with the same sample bias of -10 mV, but a different tunneling current of 50 pA in (a), 100 pA in (b), 200 pA in (c) and 400 pA in (d). The shapes of the island-like features with sharp boundaries are changing with tunneling current. The atomic Te/Se square lattice can also be resolved in (b-d). Scale bar: 2 nm. (e-h) Constant-current STM images of the same region with the same tunneling current of 200 pA, but a different sample bias of -10 mV in (e), -20 mV in (f), -30 mV in (g) and -40 mV in (h). The contrast between the island-like features and the surrounding decreases with larger negative sample bias but the shapes of the islands change little. Scale bar: 2 nm.

Figure 4

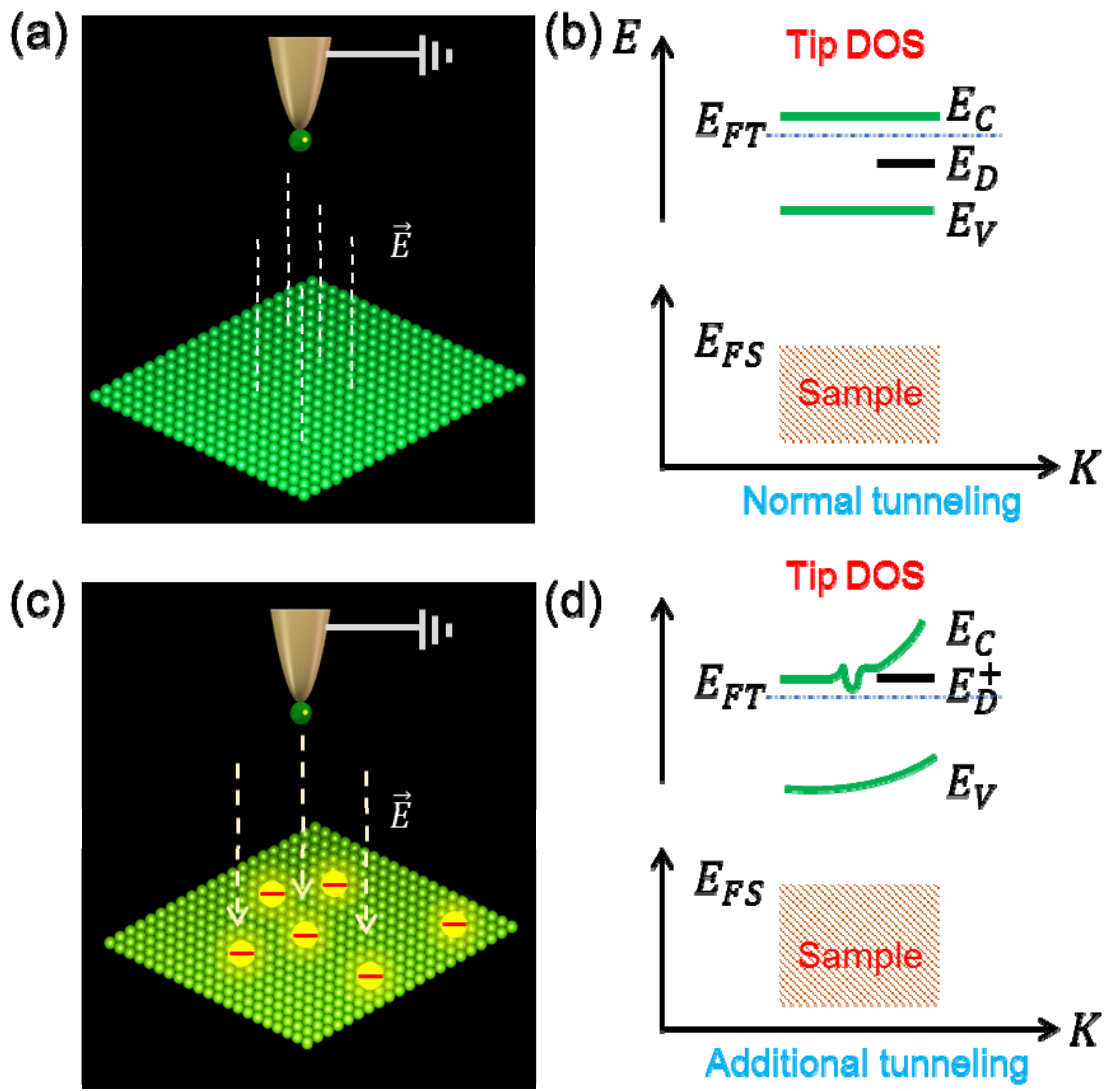

**Figure 4:** Illustration of tunneling with a semiconducting tip. (a) Schematic picture of tunneling between a semiconducting tip and normal areas of the Fe(Te,Se) sample. The tip apex is modified by a dopant atom shown as green ball. The dashed lines indicate the electric field generated by the applied sample bias voltage. (b) Illustration of the electronic bands of the tip and the sample. The Fermi surface of the tip $E_{FT}$ lies between the valence band and the conduction band. The ionic state of the dopant lies below the Fermi energy. (c) Schematic picture of tunneling between a semiconducting tip and a charged surface. An additional electric field (dashed arrows) is generated by the surface charges. (d) Illustration of the bands of the tip and sample in the presence of the additional electric field. The dopant becomes ionic by loosing its electron and the dopant ionic state is lifted above the Fermi energy. The conduction band of the tip drops because of the dopant's positive electric field and additional tunneling occurs.

# Figure 5

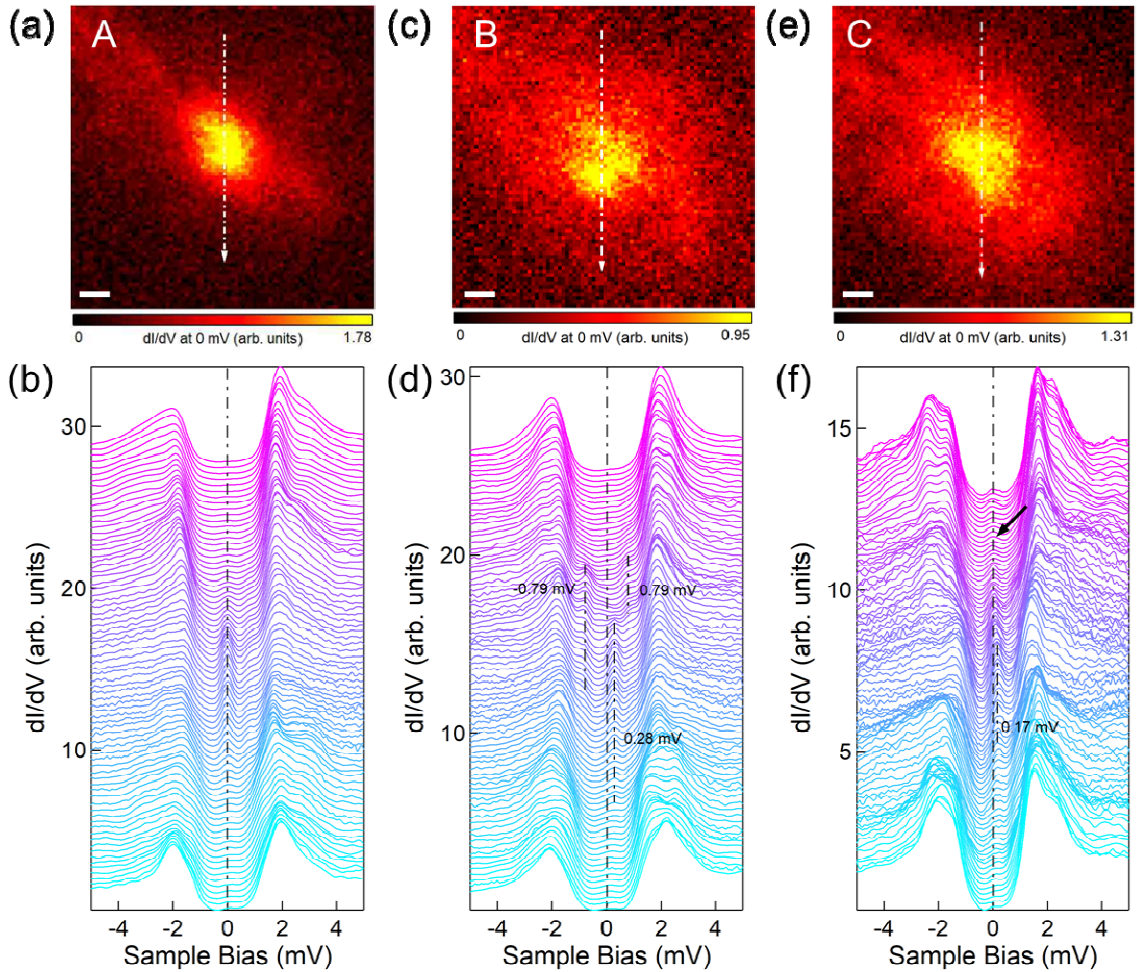

**Figure 5:** Three typical vortex bound states. (a,b) dI/dV map and STS line-cuts of vortex A. A total of 76 dI/dV spectra are presented in (b) taken along the white dashed line with a length of 15 nm shown in (a). A vortex bound state at zero energy with no spatial dispersion can be recognized. (c,d) dI/dV map and STS line-cuts of vortex B. A total of 81 dI/dV spectra are presented in (d) taken along the white dashed line with a length of 16 nm shown in (c). Within vortex B, non-dispersive CBS can be identified. (e,f) dI/dV map and STS line-cuts of vortex C. A total of 81 dI/dV spectra are presented in (f) taken along the white dashed line with a length of 17 nm shown in (e). Within the center of vortex C, the bound state has an energy of 0.17 mV. The energy of this bound state shifts to zero energy near the boundary of the vortex. Tunneling parameters in (a-f): $V_{stab}$=35mV, $I_{stab}$=100 pA, $V_{osc}$= 0.03 mV. Scale bar in (a,c,e): 2 nm.